\documentclass[preprint,showpacs,preprintnumbers,amsmath,amssymb]{revtex4}
\usepackage{graphicx}
\usepackage{dcolumn}
\usepackage{bm}

\begin{document}


\title{Coherent Dissociation of Relativistic $^9$C Nuclei}

\author{D.~O.~Krivenkov}
     \email{krivenkov@lhe.jinr.ru}
     \homepage{http://becquerel.lhe.jinr.ru}
   \affiliation{Joint Insitute for Nuclear Research, Dubna, Russia}  
\author{D.~A.~Artemenkov}
   \affiliation{Joint Insitute for Nuclear Research, Dubna, Russia}
 \author{V.~Bradnova}
   \affiliation{Joint Insitute for Nuclear Research, Dubna, Russia}
\author{M.~Haiduc}
   \affiliation{Institute of Space Sciences, Magurele, Romania}
\author{S.~P.~Kharlamov}
   \affiliation{Lebedev Institute of Physics, Russian Academy of Sciences, Moscow, Russia}
\author{V.~N.~Kondratieva}
   \affiliation{Joint Insitute for Nuclear Research, Dubna, Russia}
\author{A.~I.~Malakhov}
   \affiliation{Joint Insitute for Nuclear Research, Dubna, Russia}
\author{A.~A.~Moiseenko}
   \affiliation{Yerevan Physics Institute, Yerevan, Armenia}
\author{G.~I.~Orlova}
   \affiliation{Lebedev Institute of Physics, Russian Academy of Sciences, Moscow, Russia}
\author{N.~G.~Peresadko}
   \affiliation{Lebedev Institute of Physics, Russian Academy of Sciences, Moscow, Russia}
\author{N.~G.~Polukhina}
   \affiliation{Lebedev Institute of Physics, Russian Academy of Sciences, Moscow, Russia}
\author{P.~A.~Rukoyatkin}
   \affiliation{Joint Insitute for Nuclear Research, Dubna, Russia}
\author{V.~V.~Rusakova}
   \affiliation{Joint Insitute for Nuclear Research, Dubna, Russia}
\author{V.~R.~Sarkisyan}
   \affiliation{Yerevan Physics Institute, Yerevan, Armenia}
\author{R.~Stanoeva}
   \affiliation{Joint Insitute for Nuclear Research, Dubna, Russia}
\author{S.~Vok\'al}
   \affiliation{P. J. \u Saf\u arik University, Ko\u sice, Slovak Republic}
\author{P.~I.~Zarubin}
   \affiliation{Joint Insitute for Nuclear Research, Dubna, Russia}
 \author{I.~G.~Zarubina}
   \affiliation{Joint Insitute for Nuclear Research, Dubna, Russia}
   
\date{\today}

\begin{abstract}
\indent 
Results on the coherent dissociation of relativistic $^9$C nuclei in a nuclear track emulsion are
described. These results include the charge topology and kinematical features of final states. 
Events of $^{9}$C$\rightarrow$ 3$^{3}$He coherent dissociation are identified.\par
\indent \par
\indent DOI: 10.1134/S106377881012015X\par
\end{abstract}

\maketitle
\section{\label{sec:level1}INTRODUCTION}

\indent The coherent dissociation of relativistic nuclei on heavy target nuclei is induced in electromagnetic and nuclear diffractive interactions not accompanied by the production of target fragments and mesons. Events of this type, which are referred to as \lq\lq white\rq\rq~stars, are observed in a nuclear track emulsion with a unique reliability \cite{Baroni1,Baroni2,Andreeva}. They constitute several percent of the total number of inelastic interactions. The use of nuclear track photoemulsions ensures the completeness of observation of relativistic fragments with an excellent angular resolution. The assumption of the equality of the momenta per nucleon (or velocities) of the relativistic nucleus under study and its fragments may be a seminal hypothesis in a kinematical analysis. The angular resolution and, hence, the spatial resolution, which underlies it and which has a record value of 0.5 $\mu$m in the method of nuclear track emulsions, are of crucial importance here.\par
\indent In the dissociation of light nuclei, statistical distributions over various configurations of relativistic fragments clearly show their cluster features as a consequence of the fact that the excitation transfer is minimal \cite{Belaga,Adamovich,Shchedrina,Artemenkov,Peresadko,Stanoeva}. Investigation of neutron-deficient nuclei is particularly advantageous since the interpretation of respective experimental results ismore certain. In a nuclear track emulsion irradiated with beams from the nuclotron of the Joint Institute for Nuclear Research (JINR, Dubna), the Becquerel Collaboration \cite{web} has already studied the cluster structure of dissociation of the $^7$Be \cite{Peresadko} and $^8$B \cite{Stanoeva} nuclei. This created preconditions for proceeding to study the next isotope at the drip line, $^9$C. It can be hoped that the pattern already obtained for the $^8$B and $^7$Be nuclei must be reproduced, with an addition of one or two protons, in the coherent dissociation of the $^9$C nucleus.\par
\indent Owing to the stability of the core in the form of the $^7$Be nucleus, the $^9$C nucleus may serve as amore convenient probe of coherent-dissociation dynamics than isotopes in which the unbound nucleus $^8$Be forms a core \cite{Belaga,Shchedrina,Artemenkov}.\par
\indent In the coherent dissociation of the $^9$C nucleus, the population of the 3$^3$He cluster system, which has a relatively low formation threshold  (about 16 MeV), is possible owing to the rearrangement upon which a neutron from the  alpha-particle cluster goes over to the $^3$He cluster being formed. This system may be  of importance for the development of nuclearastrophysics scenarios as an analog  of the 3$\alpha$ process. The proposal of searches for $^{9}$C $\rightarrow$ 3$^{3}$He coherent dissociation became the main motivation of the present investigation.\par

\section{\label{sec:level2}DESCRIPTION OF THE EXPERIMENT}
\indent A secondary beam optimized to selecting $^9$C nuclei was formed via the fragmentation of $^{12}$C nuclei accelerated at the JINR nuclotron to an energy of 1.2 GeV per nucleon \cite{Rukoyatkin}. The intensity of the primary beam was about 10$^9$ nuclei per cycle, the thickness of the generating target from polyethylene was 5 g/cm$^2$, and the acceptance of the separating channel was about 3\%. This ensured the secondarybeam intensity of several hundred particles per cycle. Figure 1 shows the spectrum of the charge-to-digital converter (CDC) from the scintillation beam monitor installed in front of the track-emulsion stack used. This spectrum is indicative of a dominant contribution of carbon nuclei to the secondary beam. The beam also contains a small admixture of $^7$Be and $^8$B nuclei, which possess a somewhat higher magnetic rigidity than $^9$C. The background is due primarily to $^3$He nuclei, for which the ratio of the charge number Z$_{pr}$ to the mass number A$_{pr}$ has the same value as that for $^9$C. These features indicate that the tuning of the separating channel was correct.\par 	

\begin{figure}
    \includegraphics[width=4in]{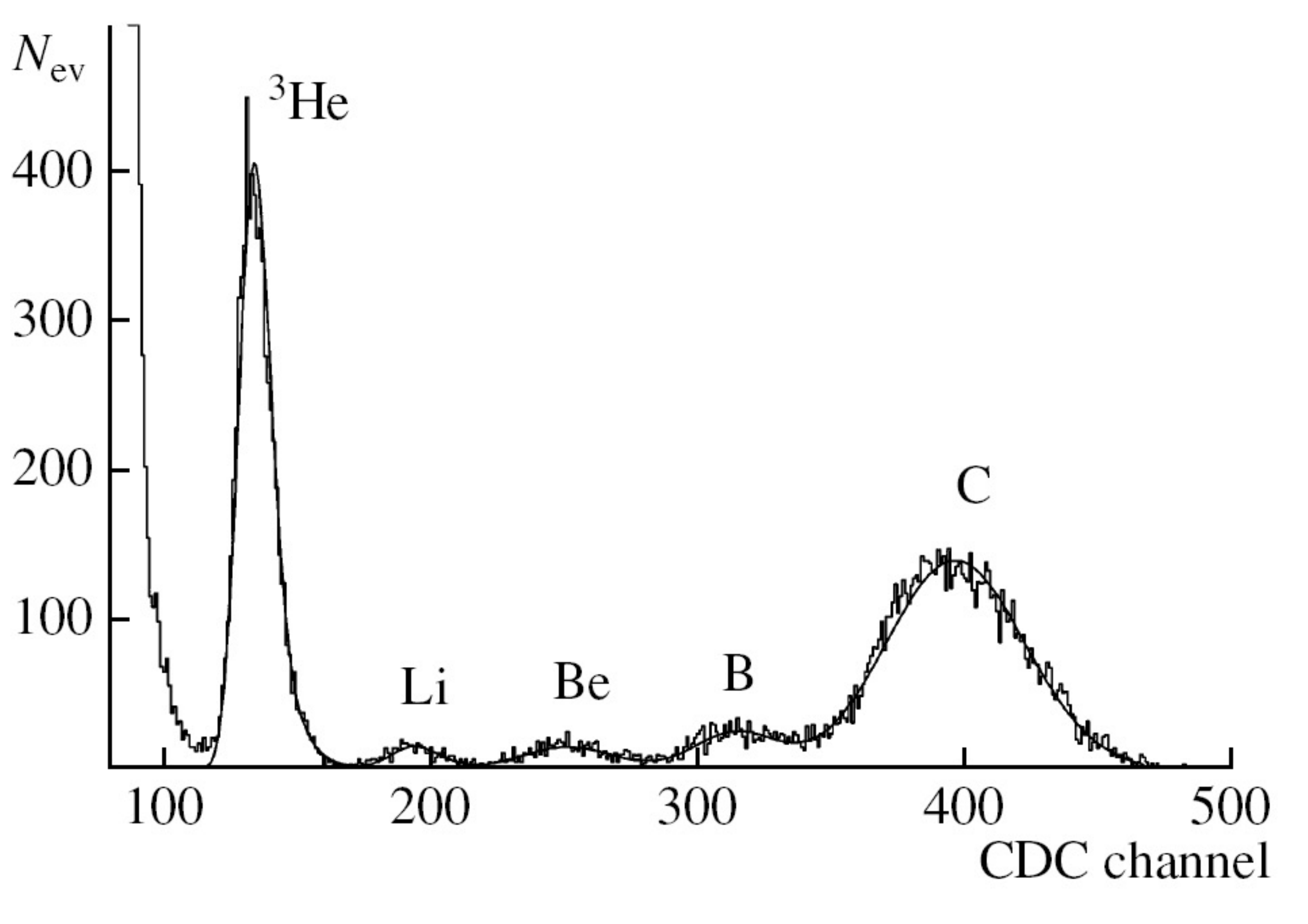}
    \caption{\label{Fig:1} Charge spectrum of nuclei from the $^{12}$C$~\rightarrow~^9$C fragmentation process in the case where the secondary beam is tuned to the value of Z$_{pr}$/A$_{pr}$ = 2/3.}
    \end{figure}

\indent The irradiated stack contained 19 layers of Br-2 nuclear track emulsion, which is sensitive even to
singly charged relativistic particles. Each layer had cross-sectional dimensions of 10 $\times$ 20 cm$^2$ and was
0.5 mm thick. During the irradiation, the beam was incident in the direction parallel to the plane of the stack along its long side with the maximum possible degree of uniformity of filling of the stack entrance window.\par
\indent The analysis described here is based on fully scanning all layers along all primary tracks for which the charges were visually estimated at values in the region Z$_{pr}~>~$2. Over the total track length of 253.7 m, we found 1746 interactions (predominantly involving carbon nuclei). We rejected $^3$He nuclei at the initial stage of visual scanning. The ratio of the intensities for Z$_{pr}~>~$2 and Z$_{pr}$ = 2 nuclei was about 0.1. This factor determined the duration of irradiation (it lasted about 100 cycles) and, as a consequence, the size of the resulting statistical sample. This irradiation had a test character and, in performing it, we tried to avoid excessively irradiating the track emulsion with $^3$He nuclei.Moreover, we discovered the presence of Z$_{pr}$ = 1 particles in approximately the same proportion as that of nuclei whose charge numbers are in excess of two, Z$_{pr}~>~$2. The mean range of carbon nuclei to interaction is $\lambda_C$ = 14.5 $\pm$ 0.5 cm, which corresponds to data for neighboring cluster nuclei.\par
\indent The identification of H and He relativistic fragments can be performed by the parameter p$\beta$c, which is determined on the basis of measuring multiple scattering; here, p is the total momentum, while $\beta$ is the velocity. We assume that projectile fragments retain a momentum per nucleon that is an integral multiple of its value for the primary nucleus; that is, p$\beta$c$~\approx$~A$_{fr}$p$_{0}\beta_{0}$c, where A$_{fr}$ is the fragment mass number. In order to attain the required accuracy in determining p$\beta$c, it is necessary to measure shifts in the coordinate of the track in directions orthogonal to the direction of particle motion at more than 100 points. The application of this laboriousmethod is justified by the derivation in this way of unique information about the isotopic composition of systems featuring several extremely light nuclei.\par

\begin{figure}
    \includegraphics[width=3in]{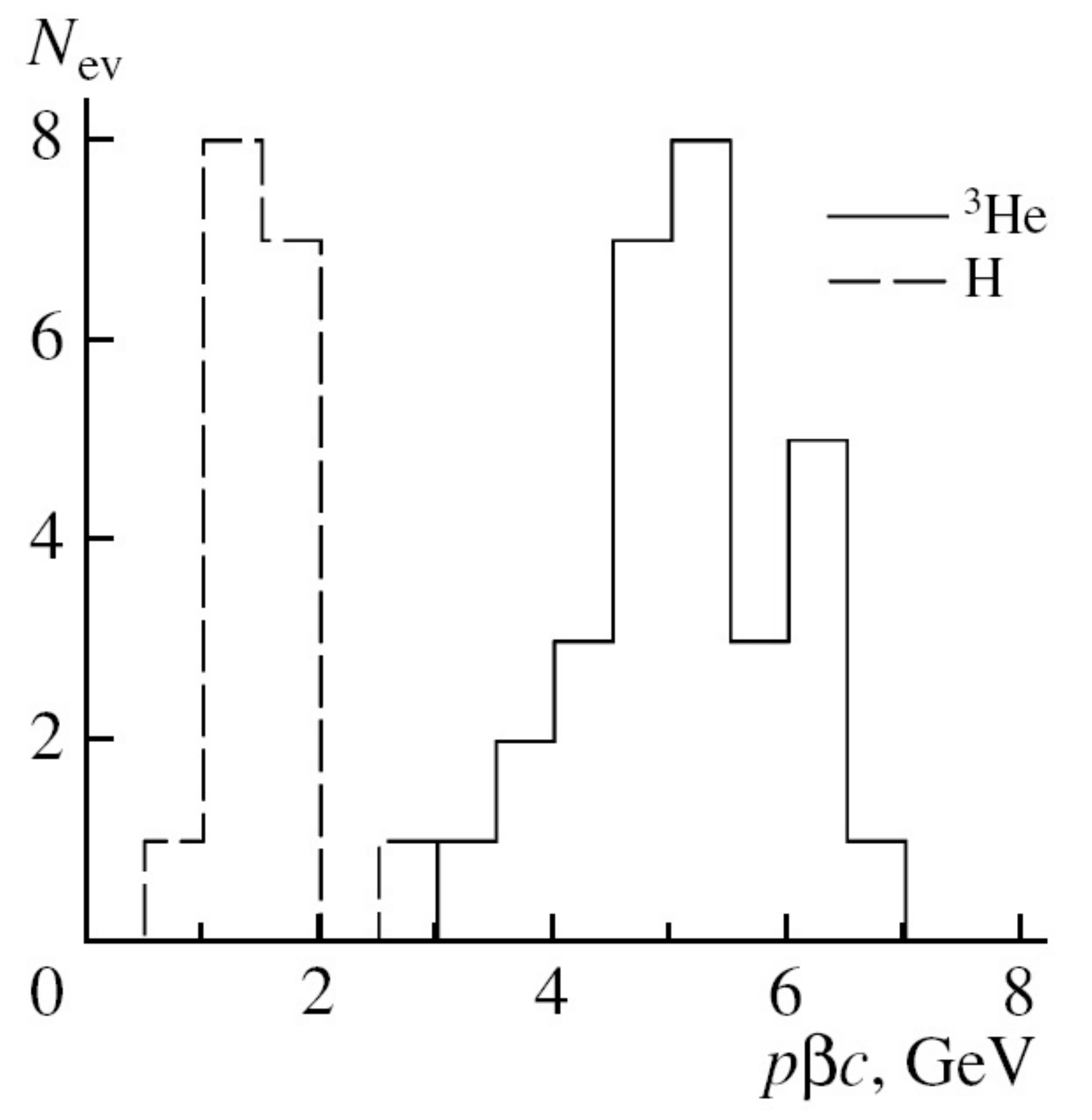}
    \caption{\label{Fig:2} Distribution of measured values of p$\beta$c for tracks of $^3$He nuclei from the beam (solid-line histogram) and singly charged fragments from $\sum$Z$_{fr}$ = 5 + 1 and 4 + 1 + 1 \lq\lq white\rq\rq~stars (dashed-line histogram).}
    \end{figure}
	
\indent The presence of $^3$He nuclei in the beam proved to be useful for calibrating the procedure for identifying secondary fragments. The distribution of measured values of p$\beta$c for 30 $^3$He nuclei from the beam composition is given in Fig. 2. The parameters of the distributions of p$\beta$c and the errors in them were determined from the results obtained by using approximations in terms of Gaussian functions. The mean value is $<$p$\beta$c$>$ = 5.1 $\pm$ 0.2 GeV, and the respective rootmean-square scatter is $\sigma$ = 0.8 GeV. Thismean value is close to that which was expected for $^3$He nuclei, 5.4 GeV (the mean value for $^4$He nuclei is 7.2 GeV). The value of $\sigma$ can be considered to be satisfactory for separating the isotopes $^3$He and $^4$He, especially within correlated groups.\par

\begin{figure}
    \includegraphics[width=3in]{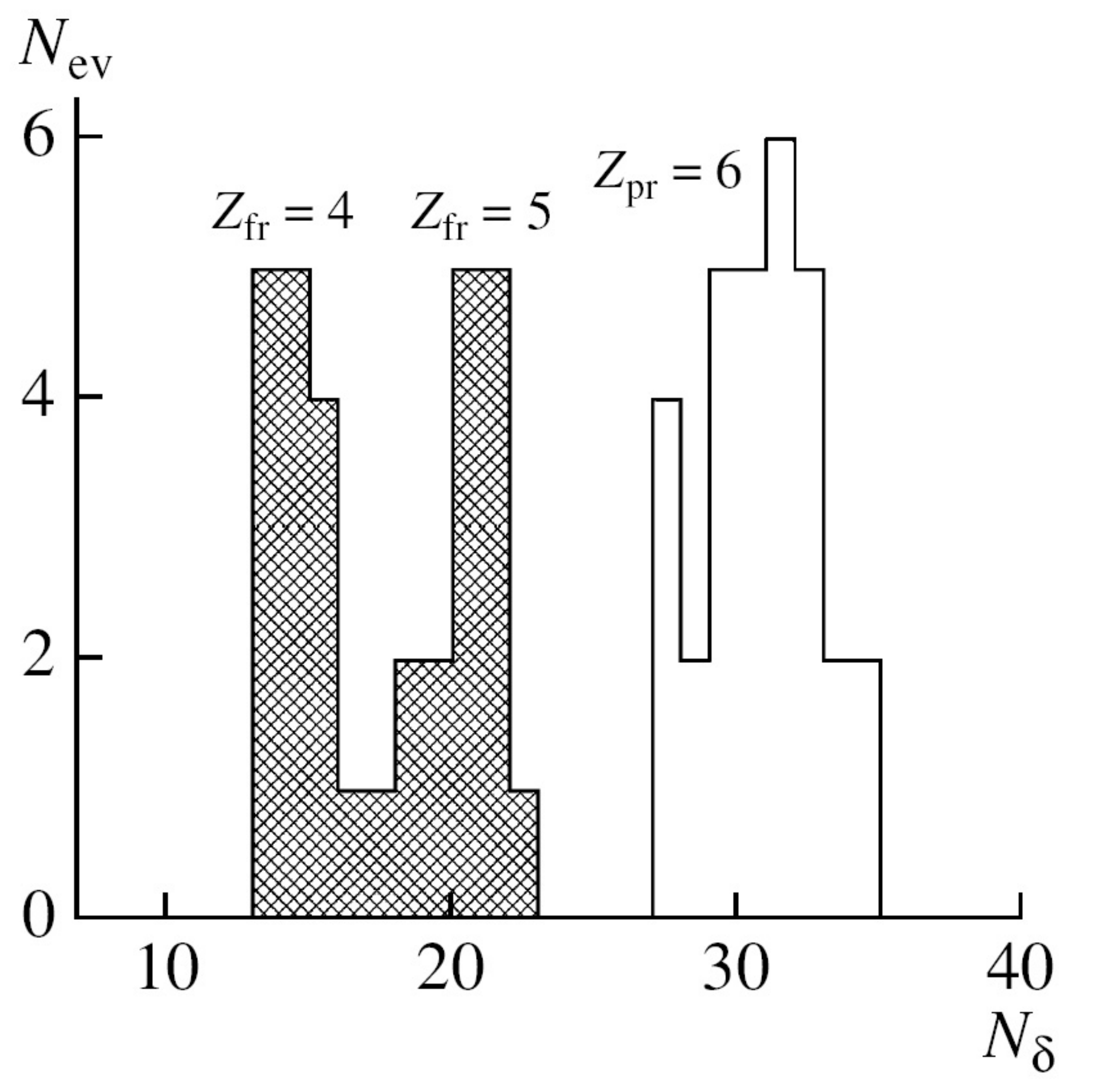}
    \caption{\label{Fig:3} Distribution of the number of events with respect to the average number of $\delta$ electrons, N$_{\delta}$, over 1 mm of length of beam-particle tracks (open histogram) and distribution of Z$_{fr}~ >~$2 relativistic fragments (shaded histogram) in $\sum$Z$_{fr}$ = 5 + 1 and 4 + 1 + 1 \lq\lq white\rq\rq~stars.}
    \end{figure}
	
\indent The charges of nuclei in the region Z$_{pr}~>~$2 were determined by considering the charge configurations of secondary fragments, $\sum$Z$_{fr}$, in \lq\lq white\rq\rq~stars and were tested by subsequently measuring the charges of primary tracks, Z$_{pr}$. The charges of beam nuclei, Z$_{pr}$, and the charges of fragments in the region Z$_{fr}~>~$2 were determined by counting $\delta$ electrons along the tracks. The results obtained by determining the charges of primary nuclei and fragments from coherent-dissociation events, $\sum$Z$_{fr}$ = 5 + 1 and 4 + 1 + 1, give sufficient grounds to conclude that all events were generated by Z$_{pr}$ = 6 nuclei (Fig. 3). For Z$_{fr}~>~$2 relativistic fragments, we can observe an expected shift of the distribution toward smaller charges in relation to the distribution for beam nuclei.\par

\section{\label{sec:level3}CONFIGURATIONS OF RELATIVISTIC FRAGMENTS}

\begin{table}
\caption{\label{Table:1} Distribution of the number of \lq\lq white\rq\rq~stars, N$_{ws}$, and the number of events involving the production of target fragments, N$_{tf}$, with respect to $\sum$Z$_{fr}$ = 6 channels}
\label{Table:1}       
\begin{tabular}{l|c|c|c|c|c|c|c|c|c}
\hline\noalign{\smallskip}
Channel & B + H & Be + 2H & 3He & Be + He & Li + He + H & Li + 3H & 2He + 2H & He + 4H & 6H \\
\noalign{\smallskip}\hline\noalign{\smallskip}
N$_{ws}$ & 15 & 16 & 16 & 4 & 2 & 2 & 24 & 28 & 6 \\
N$_{tf}$ & 51 & 47 & 9 & 7 & 11 & 8 & 54 & 80 & 16 \\
\noalign{\smallskip}\hline
\end{tabular}
\end{table}

\indent The distribution of 113 \lq\lq white\rq\rq~stars (N$_{ws}$), constituting 70\% of events of the coherent dissociation of Z$_{pr}~>~$3 nuclei, over the charge configurations $\sum$Z$_{fr}$ = 6 is given in the upper row of Table 1. Owing to the absence of the stable isotopes $^9$B and $^8$Be, events featuring fragments of charge Z$_{fr}$ = 5 and 4 and containing the identified charges of Z$_{pr}$~=~6 are interpreted as $^9$C$~\rightarrow~^8$B + $p$ and $^7$Be + 2$p$ events. These two channels have the lowest thresholds (1.3 and 1.43 MeV, respectively) and constitute about 30\% of the statistical sample of $\sum$Z$_{fr}$~=~6 coherentdissociation events. The result of identifying Z$_{fr}$~=~1 fragments from this group of events is given in Fig. 2 (dashed-line histogram). For this distribution, we have $<$p$\beta$c$>$ = 1.5 $\pm$ 0.1 GeV and $\sigma$ = 0.4 GeV, which corresponds to protons. As a matter of fact, an identification was not necessary in those cases, so that the protons in question may serve for calibration.\par
\indent Figure 4 shows the distributions of the polar emission angle $\theta$ for B, Be, and $p$ relativistic fragments of charge in the region Z$_{fr}~>~$2 within the group of events being considered. For fragments of charge Z$_{fr}$ = 5, the mean value of this angle is $<\theta_{B}>$~=~(15~$\pm$~4)~$\times$~10$^{-3}$~rad (RMS = 9.6 $\times$ 10$^{-3}$ rad, where RMS is the root-mean-square deviation), while, for fragments of charge Z$_{fr}$ = 4, it is $<\theta_{Be}>$ = (18 $\pm$ 4) $\times$ 10$^{-3}$ rad (RMS = 10.5 $\times$ 10$^{-3}$ rad).\par

\begin{figure}
    \includegraphics[width=3in]{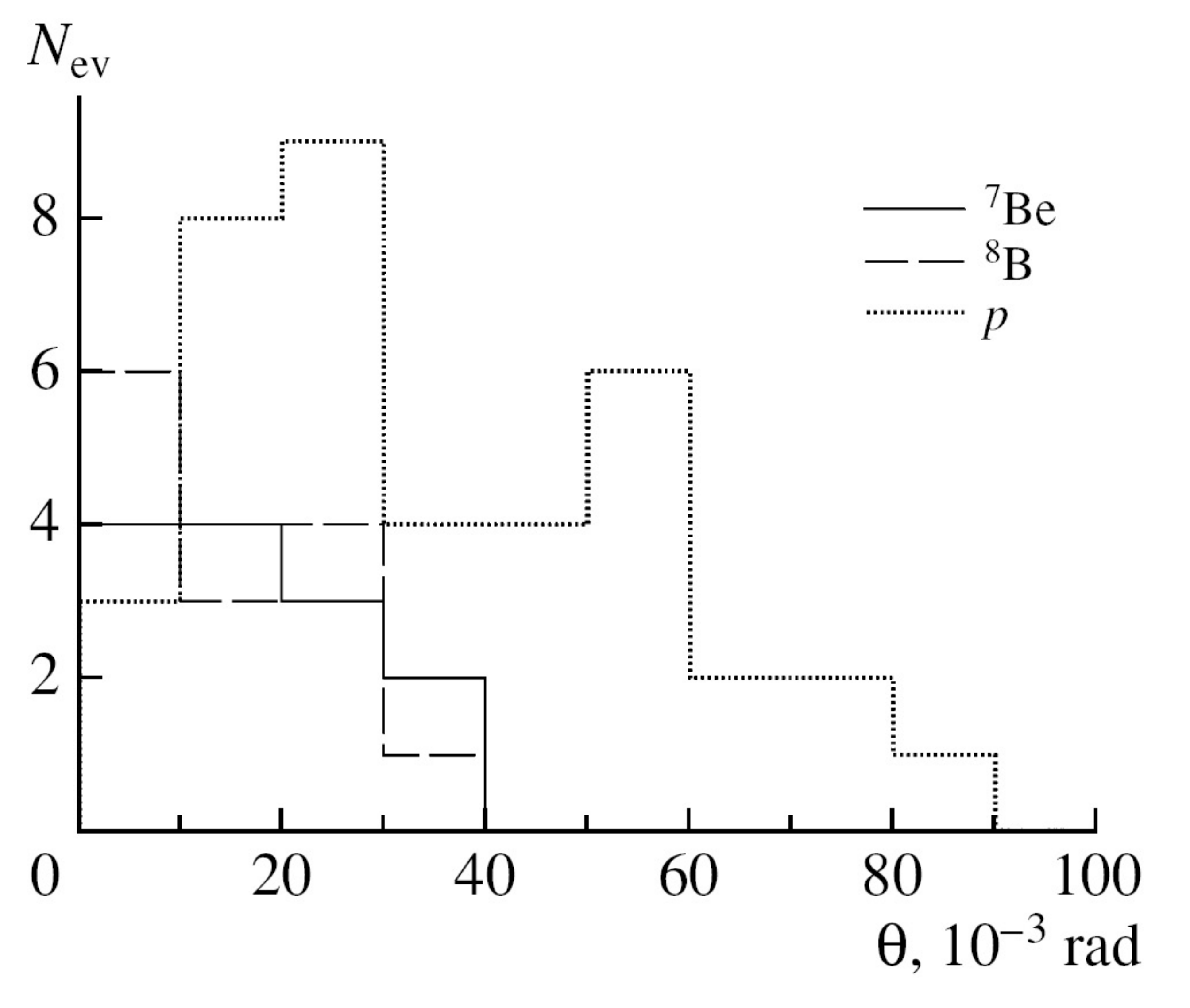}
    \caption{\label{Fig:4} Distributions with respect to the polar angle $\theta$ of relativistic fragments in the $\sum$Z$_{fr}$ = 5+1 and 4 + 1 + 1 \lq\lq white\rq\rq~stars.}
    \end{figure}

\indent The mean value $<\theta>$ for fragments of charge Z$_{fr}$~=~1 in $\sum$Z$_{fr}$~=~5~+~1 events is $<\theta_{p}>$~=~(39~$\pm$~7)~$\times$~10$^{-3}$ rad (RMS~=~26~$\times$~10$^{-3}$~rad); for $\sum$Z$_{fr}$~=~4~+~1~+~1, we have $<\theta_{p}>$~=~(34~$\pm$~4)~$\times$~10$^{-3}$~rad (RMS~=~18~$\times$~10$^{-3}$~rad). The difference in the values of $<\theta>$ for heavy and light fragments reflects the difference in their masses.\par
\indent Angular measurements make it possible to estimate, with a precision of a few percent, the fragment transverse momenta P$_t$ by using the formula \par

\indent P$_t$ = A$_{fr}$P$_0$sin$\theta$.\par

\indent The distribution of the sum of fragment transverse momenta P$_T$ reflects the mechanism of coherent dissociation. Figure 5 shows the distributions of P$_T$($^8$B~+~$p$) and P$_T$($^7$Be~+~2$p$), whose mean values are $<$P$_T$($^8$B~+~$p$)$>$~=~246~$\pm$~44~MeV/c and $<$P$_T$($^7$Be~+~2$p$)$>$~=~219~$\pm$~38~MeV/c, the RMS values being 164 and 136 MeV/c, respectively. One can conclude, in accordance with \cite{Peresadko2}, that the two distributions in question lie in a region characteristic of nuclear diffractive dissociation.\par

\begin{figure}
    \includegraphics[width=3in]{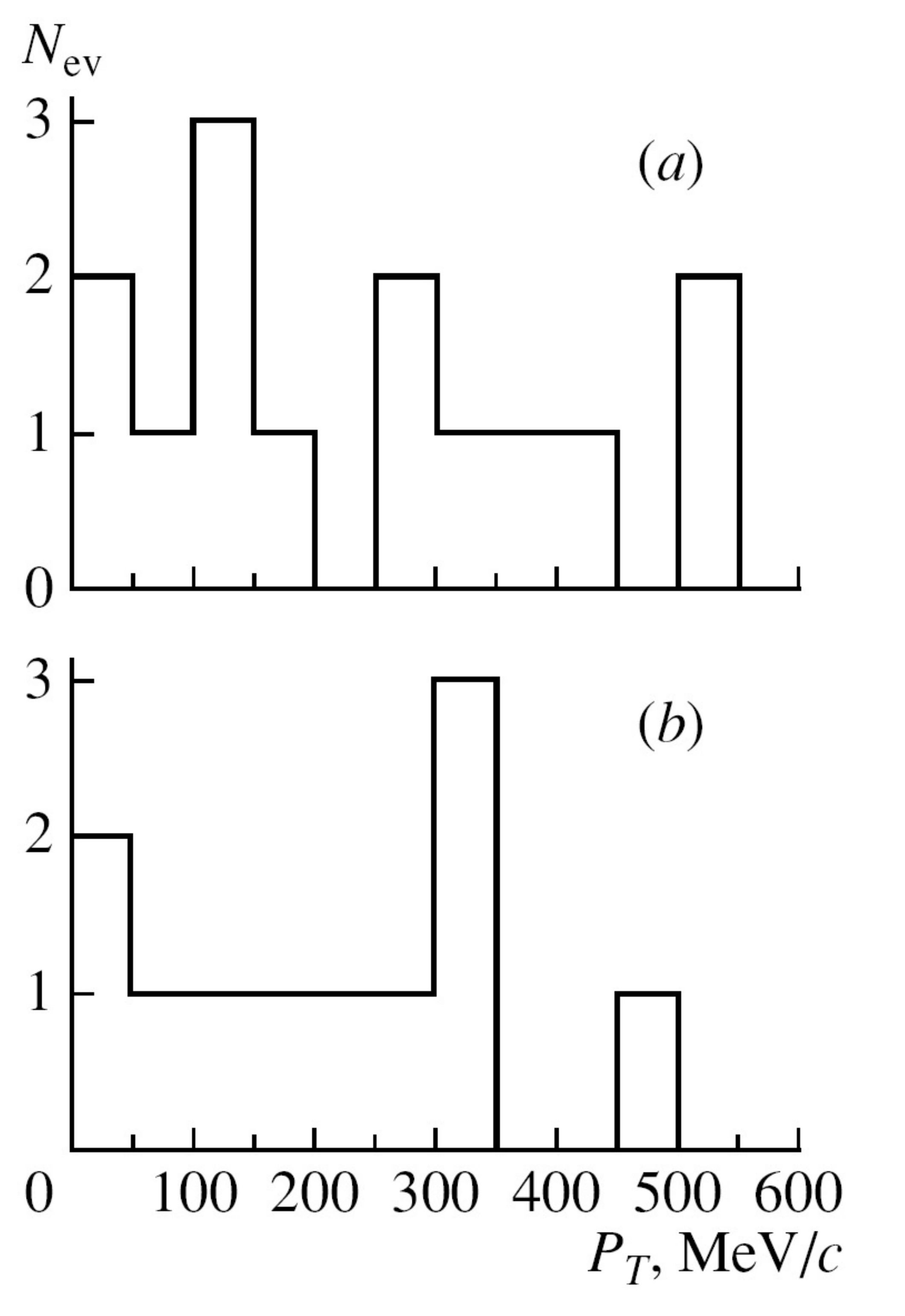}
    \caption{\label{Fig:5} Total-transverse-momentum (P$_T$) distribution of coherent-dissociation events in the reactions ($a$) $^9$C $\rightarrow$ $^8$B + $p$ and ($b$) $^9$C $\rightarrow$ $^7$Be + 2$p$.}
    \end{figure}
	
\indent The distribution of \lq\lq white\rq\rq~stars formed by $^7$Be, $^8$B, and C nuclei over the $\sum$Z$_{fr}$ charge configurations including only H and He nuclei is presented in Table 2, where one H nucleus for $^8$B and one 2H nucleus for C were excluded from $\sum$Z$_{fr}$. This group of events requires a complete identification of helium and hydrogen isotopes and a kinematical analysis, which is planned to be performed in the future. One can observe identical fractions of the 2He and He + 2H channels, and this is compatible with the dissociation of $^7$Be as the core of the $^9$C nucleus. An identification of events involving $^3$He as a product of $^9$C dissociation will be presented below. It was found in \cite{Stanoeva} that the number of $^8$B \lq\lq white\rq\rq~stars involving heavy fragments ($^8$B$~\rightarrow~^7$Be + $p$) was approximately equal to the number of stars containing only He and H fragments from the dissociation of the $^7$Be core. Under the assumption that the contribution of $^7$Be in the present study is as large as that in \cite{Stanoeva}, the statistical sample in Table 1 can be thought to correspond to the dissociation of the isotope $^9$C.\par

\begin{table}
\caption{\label{Tabel:2} Distribution of the number of \lq\lq white\rq\rq~stars formed by $^7$Be, $^8$B, and C nuclei and their relative fractions over H and He configurations}
\begin{tabular}{l|c|c||c|c||c|c}
\hline\noalign{\smallskip}
Channel & ~$^7$Be~ & ~\%~ & $^8$B(+H) & ~\%~ & $^9$C(+2H) & ~\%~ \\
\noalign{\smallskip}\hline\noalign{\smallskip}
2He & 41 & 43 & 13 & 40 & 24 & 42 \\
He+2H & 42 & 45 & 19 & 47 & 28 & 44 \\
4H & 2 & 2 & 2 & 13 & 6 & 10 \\
Li+H & 9 & 10 & 3 & 0 & 2 & 4 \\
\noalign{\smallskip}\hline
\end{tabular}
\end{table}

\indent In addition, the formation of six C $\rightarrow$ 6H \lq\lq white\rq\rq~stars is noteworthy (see Table 1). In the cases of the isotopes $^{10,11,12}$C, events of this type require a simultaneous breakup of two or three $^4$He clusters. Because of very high thresholds, they could hardly proceed without the formation of target fragments. On the contrary, similar processes associated with the breakup of only a pair of He clusters were observed for $^7$Be $\rightarrow$ 4H \cite{Peresadko} and $^8$B~$\rightarrow$~5H \lq\lq white\rq\rq~stars \cite{Stanoeva}.\par
\indent The formation of four $\sum$Z$_{fr}$ = 4 + 2 events, which could arise in the dissociation process $^{11}$C $\rightarrow$ $^7$Be + $^4$He, which has the lowest threshold for the isotope $^{11}$C, is indicated in Table 1. One can conclude that the presence of the isotope $^{11}$C in the composition of the secondary beam is insignificant. The contribution of the isotope $^{10}$C, for which configurations consisting of helium and hydrogen isotopes exclusively would be characteristic, to the statistical sample in Table 1 requires a detailed identification in this group of events. However, there are no indications of its significant role. It should be noted that, upon such an identification, new physical conclusions can be drawn, since cases corresponding to the intersection of the drip line in the direction of nuclear resonance states can be discovered in the fragmentation process $^9$C~$\rightarrow$~$^8$C.\par
\indent For the sake of comparison, the number of events accompanied by target fragments for $\sum$Z$_{fr}$ = 6, N$_{tf}$, is presented in the lowest row of Table 1. In almost all of the channels, there is an approximate proportionality to N$_{ws}$. This is not so only for $^3$He, whose yield N$_{tf}$ exhibits a pronounced decrease.\par

\section{\label{sec:level3}SEARCH FOR EVENTS INVOLVING 3$^3$He}
\indent The formation of 16 \lq\lq white\rq\rq~stars having 3He final states is indicated in Table 1. This made it possible to address the problem of identifying $^9$C $\rightarrow$ 3$^3$He events. It is worth noting that the probability of dissociation through this channel is commensurate with the probability for the most expected final states. Measurements of angles make it possible to obtain distributions with respect to the polar angle of fragment emission, $\theta$ (see Fig. 6), and with respect to the pair opening angle, $\Theta_{2He}$ (see Fig. 7). Eight narrow 2He pairs for which $\Theta_{2He}~<~10^{-2}$ rad were reliably observed owing to an excellent spatial resolution.\par

\begin{figure}
    \includegraphics[width=3in]{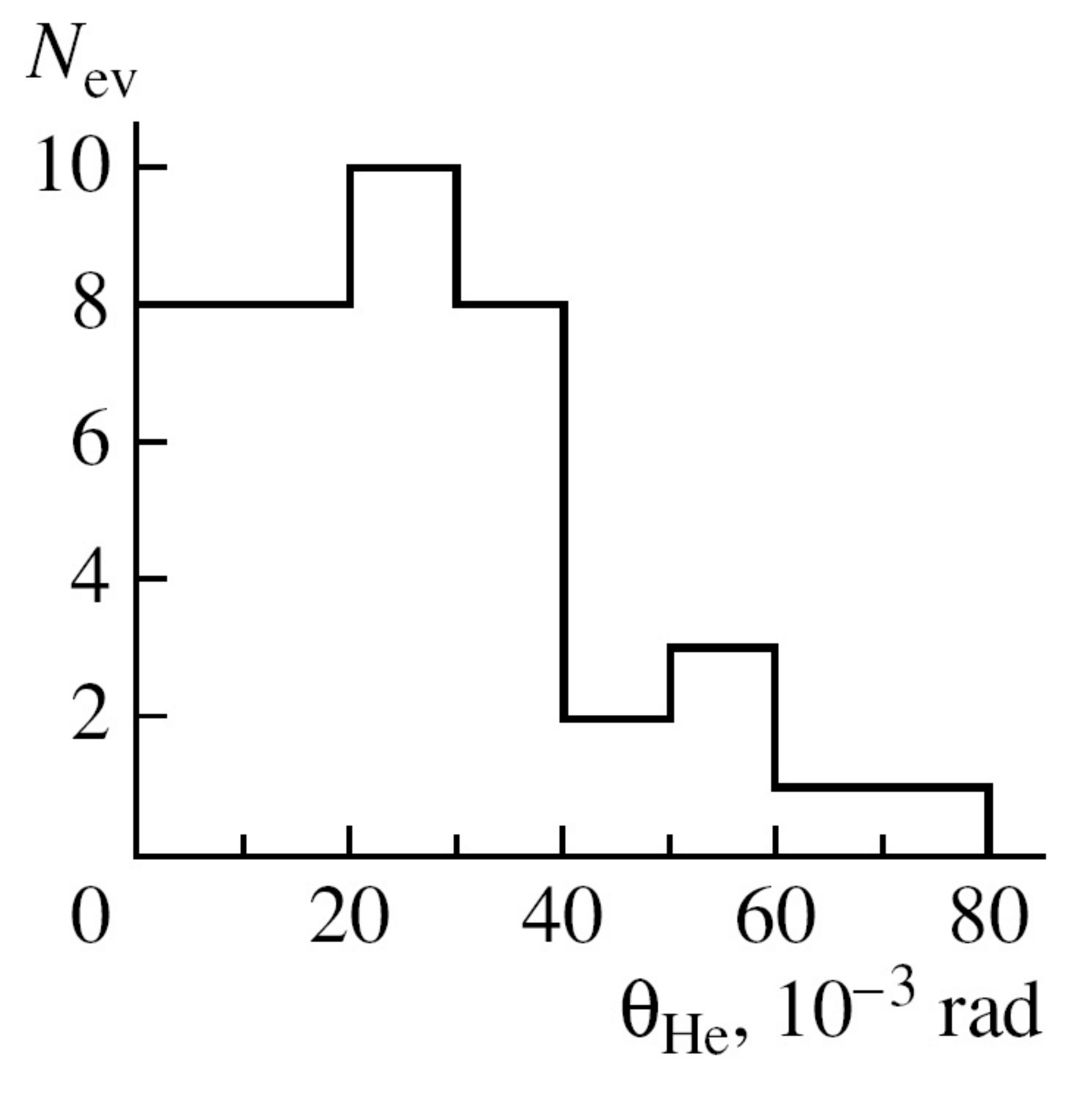}
    \caption{\label{Fig:6} Distribution of the polar angle $\theta$ for doubly charged fragments in C $\rightarrow$ 3He \lq\lq white\rq\rq~stars.}
    \end{figure}
	
\begin{figure}
    \includegraphics[width=3in]{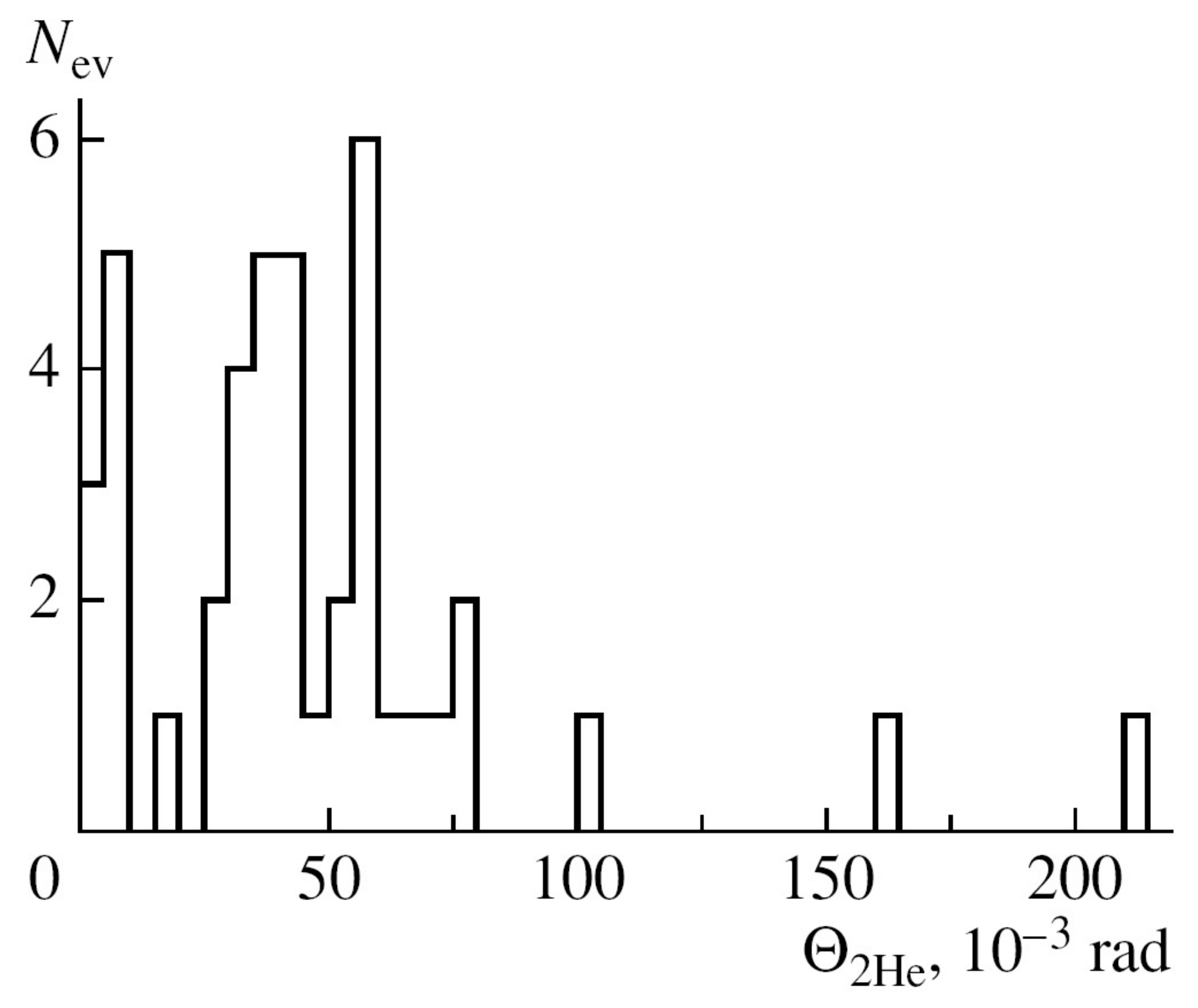}
    \caption{\label{Fig:7} Distribution of the angle $\Theta_{2He}$ between the fragment momenta in C $\rightarrow$ 3He \lq\lq white\rq\rq~stars.}
    \end{figure}\
	
\indent This channel could be identified by three He fragments. However, the situation actually prevalent during
the exposure of the track emulsion to the secondary beam turns out to be more intricate. The admixture of $^{10}$C nuclei could also lead to events of a deep rearrangement of nucleons in the process $^{10}$C $\rightarrow$ 2$^3$He + $^4$He. In order to identify helium isotopes, we therefore employed measurements of multiple scattering. Such measurements could be performed only for 22 tracks (Fig. 8). We obtained the mean value of $<$p$\beta$c$>$~=~4.9~$\pm$~0.3~GeV and the root-mean-square deviation of $\sigma$~=~0.9 GeV. This corresponds to the results of the calibration performed by using nuclei of a $^3$He beam. The fraction of fragments that could be identified as $^4$He nuclei is insignificant in relation to the fraction of $^3$He.\par

\begin{figure}
    \includegraphics[width=4in]{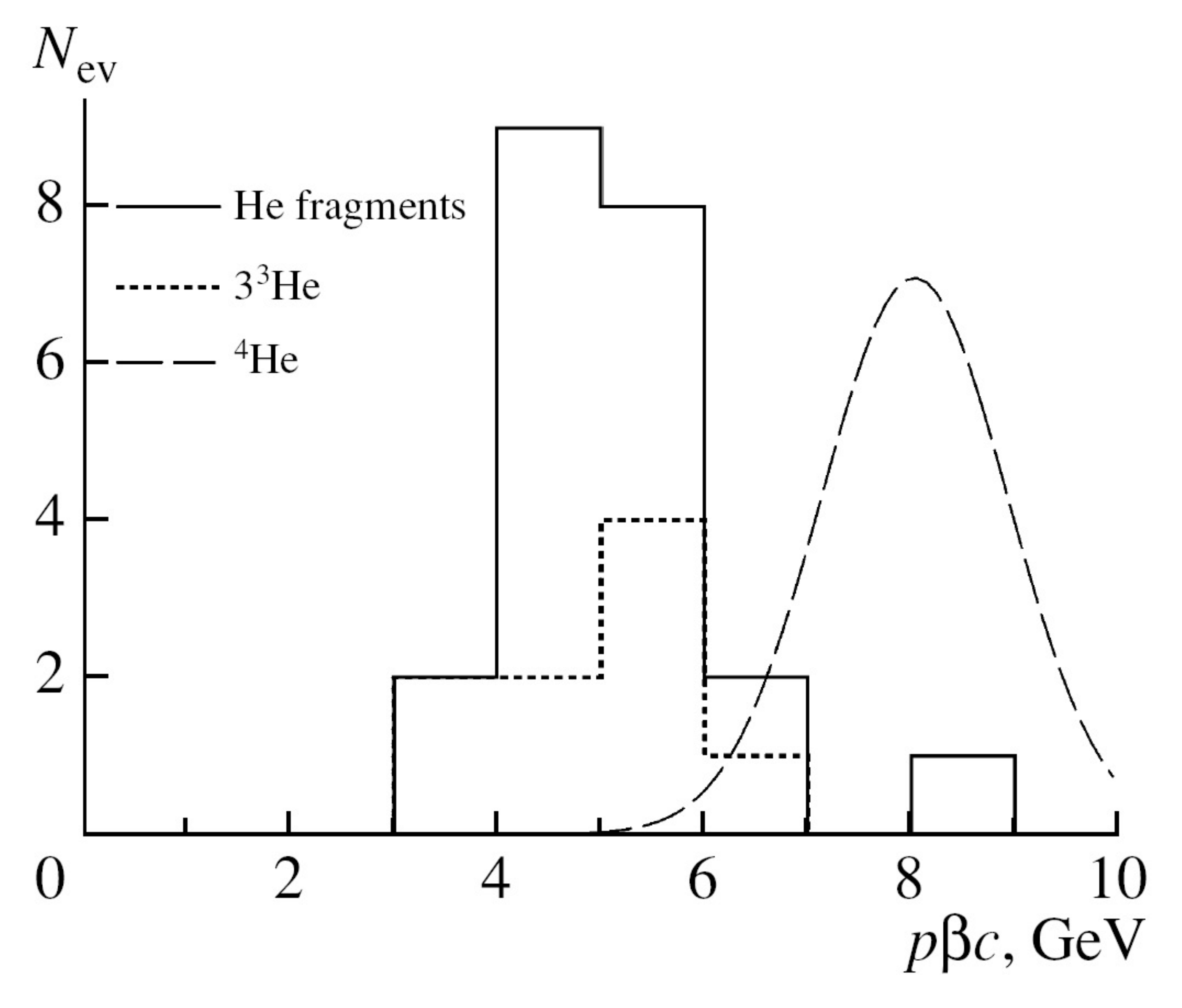}
    \caption{\label{Fig:8} Distribution of measured values of p$\beta$c for doubly charged fragments from (solid-line histogram) $^3$He \lq\lq white\rq\rq~stars and (dotted-line histogram) fully identified 3$^3$He events. The dashed curve represents the distribution expected for $^4$He.}
    \end{figure}
	
\indent A determination of p$\beta$c for all fragments in 3He events could be performed in only three of such events (see Fig. 8). The resulting values make it possible to interpret these events as the triple production of $^3$He nuclei. The interpretation of these events as those of the reaction $^{10}$C $\rightarrow$ 3$^3$He + $n$ is improbable, since, in that case, the requirement of a peripheral interaction without the formation of target fragments would entail a modification of a pair of $^4$He clusters (rather than one such cluster), in which case it is necessary to overcome a threshold of at least 37 MeV. The microphotograph of one identified $^9$C $\rightarrow$ 3$^3$He event is shown in Fig. 9\par

\begin{figure}
    \includegraphics[width=6in]{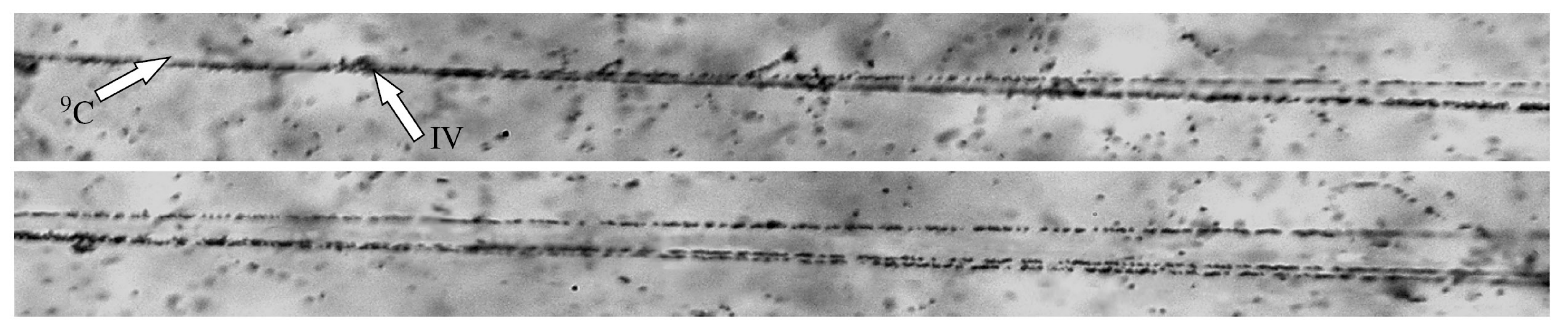}
    \caption{\label{Fig:9} Microphotograph of a $^9$C $\rightarrow$ 3$^3$He \lq\lq white\rq\rq~star at an energy of 1.2 GeV per nucleon. The upper photograph shows the dissociation vertex and a fragment jet in a narrow cone; moving along the jet, one can see three He relativistic fragments (lower photograph).}
    \end{figure}
	
\indent As in the case presented in Fig. 5, the distribution of the total transverse momentum for the $^3$He configuration (see Fig. 10) has a form characteristic of nuclear diffractive dissociation \cite{Peresadko2}. Its parameters have somewhat larger values: $<$P$_T$(3$^3$He)$>$ = 335 $\pm$ 79 MeV/c, the RMS value being 294 MeV/c. The values of the parameters in question in fully identified events of the $^9$C $\rightarrow$ 3$^3$He transition are compatible with those values, but their accuracy is substantially poorer.\par

\begin{figure}
    \includegraphics[width=3in]{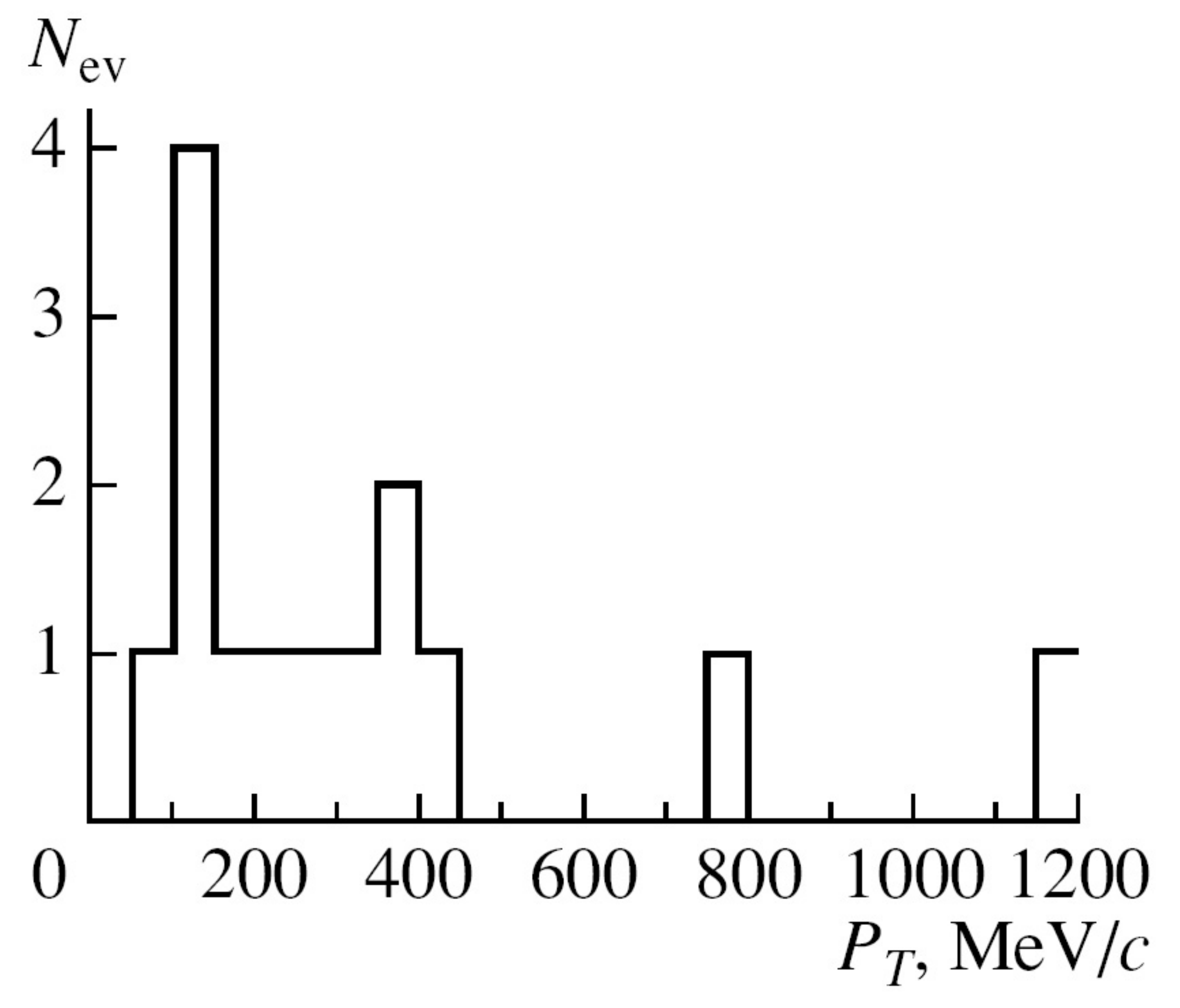}
    \caption{\label{Fig:10} Total-transverse-momentum (P$_T$) distribution of events of $^9$C $\rightarrow$ 3$^3$He coherent dissociation.}
 \end{figure}

\section{\label{sec:level4}CONCLUSIONS}	
\indent Special features of nucleon clustering in the coherent dissociation of $^9$C relativistic nuclei have been studied for the first time by the method of nuclear track emulsions. By and large, the results comply with observations for neighboring neutron-deficient nuclei, the diffractive nuclear interaction determining the dynamics of the coherent dissociation of $^9$C nuclei.\par
\indent It is noteworthy that the high-threshold $^9$C $\rightarrow$ 3$^3$He channel has a significant weight commensurate with weights of low-energy channels involving the separation of one or two nucleons. Possibly, this circumstance is indicative of a sizable admixture of the 3$^3$He virtual component in the structure of the ground state of the $^9$C nucleus. In that case, this component must contribute to the $^9$C nuclear magnetic moment, which has an anomalous value from the point of view of the shell model \cite{Utsuno}.\par
\indent In our investigation of the 2$\alpha$ + $n$ structure of the $^9$Be nucleus \cite{Artemenkov}, it was found that the $^9$Be $\rightarrow$ 2$\alpha$ fragmentation process proceeds with close probabilities through the 0$^+$ and 2$^+$ states of the $^8$Be nucleus. These probabilities show correspondence with the weights assigned to the 0$^+$ and 2$^+$ states of the $^8$Be nucleus in calculating the magnetic moment of the $^9$Be nucleus on the basis of cluster wave functions \cite{Parfenova1,Parfenova2}. Data on the relativistic fragmentation of $^9$Be nuclei \cite{Artemenkov} can be considered as the proof that, with a high probability, the ground-state structure of the $^9$Be nucleus is formed by a $^8$Be nucleus, appearing in two states and playing the role of a core, and an outer neutron. Thereby, there arise new arguments in support of considering data on final states of relativistic fragmentation as a reflection of weights of cluster components of the ground states of the nuclei under study.\par
\indent The development of this conclusion for the $^9$C nucleus deserves verification via calculating themagnetic moment on the basis of cluster wave functions along the same lines of reasoning as those that were used in the analysis performed for the $^9$Be nucleus in \cite{Parfenova1,Parfenova2}.\par
\begin{acknowledgments}
\indent This work was supported by the Russian Foundation for Basic Research (project nos. 96-1596423, 02-02-164-12a, 03-02-16134, 03-02-17079, 04-02-17151, 04-02-16593, and 09-02-9126-ST-a) and by the Agency for Science at theMinistry for Education of the Slovak Republic and Slovak Academy of Sciences (grants VEGA nos. 1/2007/05 and 1/0080/08), as well as by grants from the Plenipotentiaries of Bulgaria, the Slovak Republic, the Czech Republic, and Romania at the Joint Institute for Nuclear Research (JINR, Dubna) in 2002–2009. We gratefully acknowledge the meticulous work of A.M. Sosul'nikova (JINR) on visually scanning track emulsions.\par
\indent \par
\indent Translated by A. Isaakyan \par
\end{acknowledgments}

\end{document}